\documentclass[12pt]{article}

\usepackage{scicite}

\usepackage{times}

\usepackage{amsmath}
\usepackage{graphicx}
\usepackage{comment}
\usepackage{booktabs}
\usepackage{threeparttable}
\usepackage{multirow}
\usepackage{mdwtab}

\usepackage{caption}

\newenvironment{sciabstract}{%
\begin{quote} \bf}
{\end{quote}}

\title{Deconstructing laws of accessibility and facility distribution in cities}

\author
{Yanyan Xu,$^{1,2,3}$ Luis E. Olmos,$^{1,2,3}$ Sofiane Abbar,$^{4}$ Marta C. Gonz\'{a}lez$^{1,2,3,5\ast}$\\
\\
\normalsize{$^{1}$Department of Civil \& Environmental Engineering, Massachusetts Institute of Technology,}\\
\normalsize{Cambridge, MA 02139, USA}\\
\normalsize{$^{2}$Department of City and Regional Planning, University of California,}\\
\normalsize{Berkeley, CA 94720, USA}\\
\normalsize{$^{3}$Energy Technologies Area, Lawrence Berkeley National Laboratory,}\\
\normalsize{Berkeley, CA 94720, USA}\\
\normalsize{$^{4}$Qatar Computing Research Institute, HBKU, Doha 5825, Qatar}\\
\normalsize{$^{5}$Department of Civil and Environmental Engineering, University of California,}\\
\normalsize{Berkeley, CA 94720, USA}\\
\\
\normalsize{$^\ast$To whom correspondence should be addressed. E-mail: martag@berkeley.edu.}
}

\date{}

\begin{document} 

\captionsetup[figure]{labelfont={bf},labelformat={default},labelsep=period,name={Fig.}}


\baselineskip24pt


\maketitle


\begin{sciabstract}
The era of the automobile has seriously degraded the quality of urban life through costly travel and visible environmental effects. A new urban planning paradigm must be at the heart of our roadmap for the years to come. The one where, within minutes, inhabitants can access their basic living needs by bike or by foot. In this work, we present novel insights of the interplay between the distributions of facilities and population that maximize accessibility over the existing road networks. Results in six cities reveal that travel costs could be reduced in half through redistributing facilities. In the optimal scenario, the average travel distance can be modeled as a functional form of the number of facilities and the population density. As an application of this finding, it is possible to estimate the number of facilities needed for reaching a desired average travel distance given the population distribution in a city.
\end{sciabstract}

\section*{Introduction}

At a time of the very visible effects of the climate impact on our urban lives, some cities have become unbreathable, and greenhouse gas emissions are produced by buildings heating and cooling networks, and all-round petrol transport. At a time when transport has become the first emitter of CO$_{2}$, we need to imagine, propose, other ways of occupying urban space. This calls for a better understanding of the spatial distributions of facilities and population~\cite{makse1995modelling,pan2013urban,bettencourt2007growth,bettencourt2010unified, glaeser2011cities,bettencourt2013origins,weiss2018global}. The information age and the online mapping revolution allow us to globally study the interactions of humans with their built and natural environment~\cite{rozenfeld2011area,tsekeris2013city,bettencourt2014professional,schlapfer2014scaling,brelsford2017heterogeneity,li2017simple}. Pioneering work in multi-city studies have uncovered scaling laws relating population to distribution of facilities and socio-economic activities at macroscopic scale~\cite{newman2005power,bettencourt2007growth,bettencourt2013origins,depersin2018global,barthelemy2019statistical}. It has been asserted, for example, that more populated cities are more efficient in their per capita consumption~\cite{bettencourt2007growth,bettencourt2010unified} and their occupation diversity can be modeled as social networks embedded in space~\cite{bettencourt2014professional}. Yet, a systematic understanding of the interplay of the urban form, their facilities distribution and their accessibility at multiple scales remains an elusive task.

At the country scale, when maximizing for the accessibility of population to a fixed number of facilities, Gastner and Newman demonstrated a simple $2/3$ power law between the optimal density of facilities $d$ and their population density $\rho$~\cite{gastner2006optimal}. The power law was fitted by allocating $5,000$ facilities in the continental U.S. using population data within more than 8 million census blocks. In this case, each facility covers an area about the size of a county ($\sim$1,000 km$^{2}$). In a follow up study, Um et~al. proposed distinct optimization goals to differentiate public services, such as fire stations and public schools, from commercial facilities, such as banks and restaurants~\cite{um2009scaling}. Public service facilities aim to minimize the overall distance between people and the facilities, follow $d \propto \rho^{2/3}$. However, in the case of profit driven facilities, which have the goal of maximizing the number of potential customers, the power law has an exponent close to 1, that is $d \propto \rho$. The authors found alignment in the analytical optimization and empirical distributions in the U.S. and South Korea, confirming the $2/3$ exponent for public services and the $1$ exponent for profit driven facilities. The simple power law at city scale reveals the equilibrium of empirical allocation of resources across cities with different population. However, distributing facilities at fine scale within cities, where the coverage area per facility is of few blocks ($\sim$10 km$^{2}$), results in more heterogeneous settlements of population with different socio-economic characteristics. Studies of accessibility within cities merit attention for science informed land use planning and the redistribution of public services after disasters and evacuations~\cite{tsou2005accessibility,apparicio2006measuring,macintyre2008poorer,dadashpoor2016inequality,brelsford2018toward}. Forward-looking approaches for planning facilities in cities would also consider individuals’ preferences to facilities via mining mobility patterns. Zhou et~al. introduced a location-based social network dataset to derive the demand for different types of cultural resources, and identified the urban regions with lack of venues~\cite{zhou2018discovering}. While efforts have been devoted to address the optimal allocation problem in specific cities~\cite{white1979accessibility,tao2014spatial,zhou2018discovering,olmos2020data}, systematic understanding of the optimal distribution of facilities is still lacking from the urban science perspective.

To contribute in this direction, we propose a multi-city study that measures the accessibility of city blocks to different types of facilities through their road networks, and investigate the role of population distributions. While at large scale, travel cost can be substituted by the Euclidean distance from residents to the facilities, road networks and geographic constraints play important roles for human mobility within cities~\cite{samaniego2008cities,louf2014congestion,louail2014mobile,florez2017measuring}. It has been well established that road network properties impact the daily journeys of residents~\cite{louf2013emergence,wang2014encapsulating,ccolak2016understanding}, their urban form~\cite{strano2012elementary,louf2014typology} and their accessibility~\cite{louf2013modeling,louf2014congestion}. As a complement to most studies devoted to travel costs of commuters, we analyze in this work the road network distance of individuals to the nearest amenity of various types, dividing the space in high resolution blocks of constant area of 1 km$^2$. For each city and facility type, we optimally redistribute the existing facilities and compare the result with their empirical distribution. We observe that in the redistribution some blocks increase their accessibility and others decrease it. This implies that in order to make the best use of the existing facilities for a more equitable accessibility, some blocks would benefit whereas others would have facilities removed. At the city level, the gap between the empirical facility distribution and the optimal planning offers the opportunity to assess the planning quality of facilities in diverse cities. We also revisit the power law between facility and population densities, and observe that the two-thirds power law is not followed by the empirical cases, and it is observed in the optimal scenario only when the number of facilities is small compared to the total number of blocks in the city.

We further investigate optimal distributions of facilities by modeling its average travel distance in different cities as a function of the number of facilities to assign. A model of this quantity is derived on both synthetic and real-world cities and fits different cities well with only two free parameters. Furthermore, this gives us a universal function between the average travel distance and the number of facilities, controlled by the urban form derived from the population distribution. As an application case, we estimate the number of facilities required to achieve a given accessibility via the proposed function in $12$ real-world cities.

\section*{Results}

\subsection*{Empirical Distribution of Facilities}

We select three cities (Boston, Los Angeles, New York City) in the U.S. and three cities (Doha, Dubai, Riyadh) in the Gulf Cooperation Countries (GCC) to study the empirical distribution of facilities. For each city we collect the population in blocks with a spatial resolution of 30 arc-seconds (1 km$^2$ near the equator) from LandScan~\cite{landscan2015}, road networks with the OpenStreetMap~\cite{osm2017}, and facilities from the Foursquare~\cite{foursquare2017} service application. These novel, rich, and publicly available datasets have proven value in transportation planning~\cite{toole2015path,ccolak2016understanding,xu2017collective}, land use studies~\cite{quercia2014mining,spyratos2017using}, and human activity modeling~\cite{frith2014communicating,krueger2015semantic,xu2019unraveling}. The boundary of each city is drew along with the metroplex, encompassing both urban and rural regions. Fig.~1 depicts the road network, population density, and ten selected types of facilities (e.g., hospitals, schools) in New York City (NYC) and Doha. The statistical information of six cities are summarized in Table~1. For clarity, all variables and notations introduced in this work are summarized in note~S1. The distribution of all available facilities in Foursquare data for the six cities are presented in fig.~S1. Details of the data sets are described in the \textit{Materials and Methods} and note~S2. Fig.~S2 presents the distribution of population and different facility categories as a function of the distance from the central business district (CBD), indicating the diversity of the selected cities. Specifically, it can be observed that Doha and Dubai have more facilities that are located in the highly populated areas, whereas Boston has the majority of facilities located near the city center with fewer people reside. Discrepancy in the distributions between population and facilities can also be observed in Los Angeles (LA), NYC and Riyadh. In these three cities, the population density peaks near the city center, but the facilities are distributed more uniformly across the city. 

\clearpage

\begin{figure*}[htb!]
\centering
\includegraphics[width=0.90\linewidth]{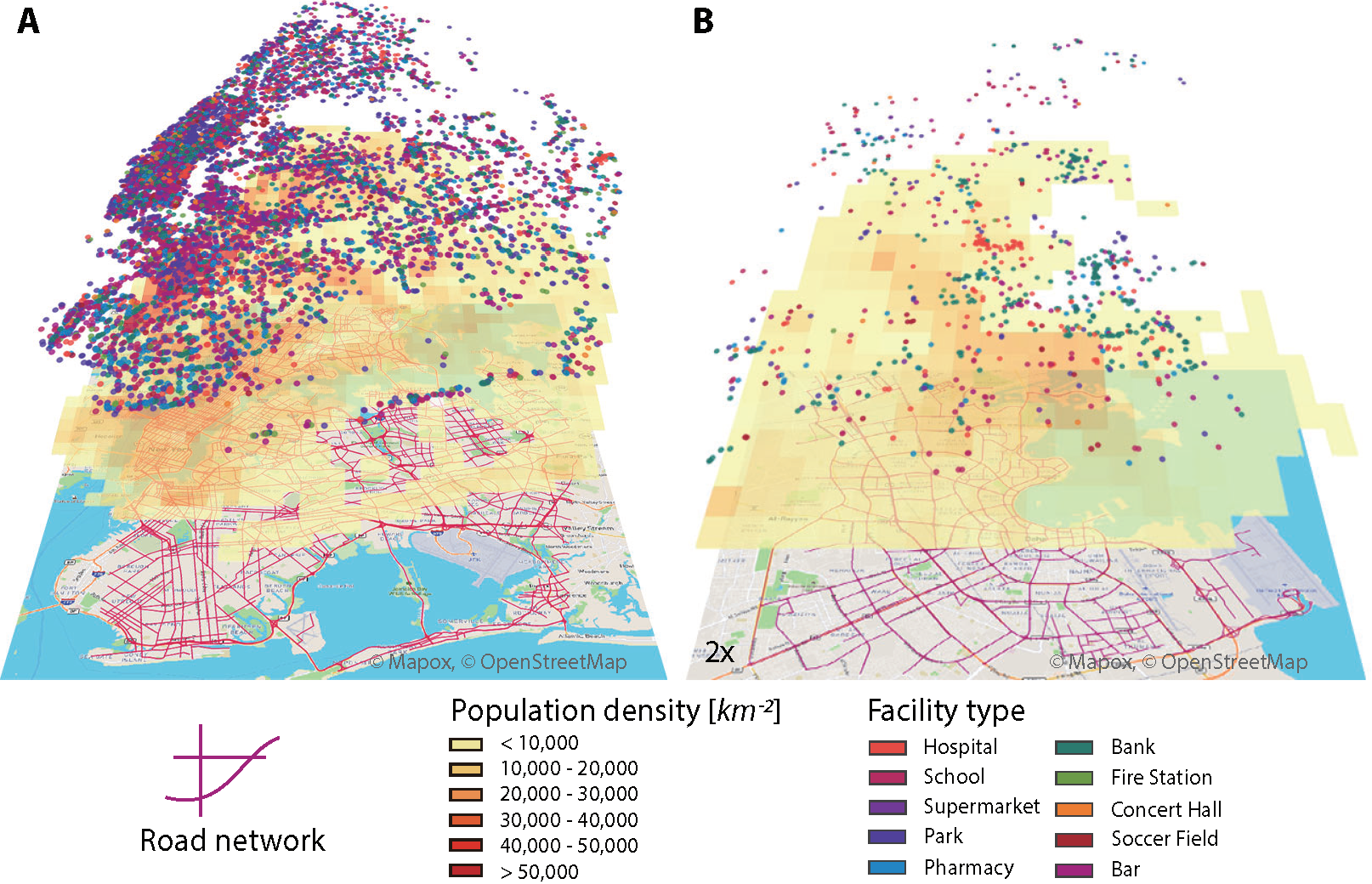}
\caption{{\bf Illustration of the three datasets in NYC (\textbf{A}) and Doha (\textbf{B}).} The lower layer depicts the map and road networks of the city. The middle and upper layers illustrate the population density and the locations of the ten selected types of facilities in the same region, respectively. Compared with very dense NYC, Doha has a simpler road network, less population and sparser facilities.}
\end{figure*}

It is noteworthy that, for calculating facility density and total number of facilities, we first merge the same type of facilities (e.g., hospitals) located in the same block as one facility. Thus, the number of facilities thereafter refers to the number of blocks accommodating a given type of facility, denoted by $N$. We define $N_{max}$ as the total number of blocks in one city. Furthermore, as nearly unpopulated blocks do not weight in the calculations of accessibility, we define $N_{occ}$ as the number of occupied blocks given by the blocks with population over a threshold. We set the threshold as $500$ in real-world cities, which is commonly used to distinguish between urban and rural regions. The ratio between the number of blocks occupied by facilities $N$ and populated blocks $N_{occ}$ is denoted by $D_{occ}$. Table~1 reports the $D_{occ}$ of the ten selected types of facilities in the six cities of study. As an example, $D_{occ}$ of hospital in Boston equals $0.11$ indicating that about $11\%$ of populated blocks are occupied by hospitals.

To quantify the accessibility of the population to facilities, previous work used the Voronoi cell around each facility, as a proxy of the tendency of individuals to select the closest facility in Euclidean distance~\cite{gastner2006optimal,um2009scaling}. However, within cities, the distance that people travel in the road networks is constrained by the infrastructure and the landscape. In this context, the routing distance is a better proxy of the accessibility from the place of residence to each amenity. Fig.~S2C compares the distributions of routing distance of the actual and optimal locations of facilities vs. the Euclidean distances, respectively. Interestingly, our findings confirm that the optimal strategy based on Euclidean distance achieves similar costs to the actual distribution of facilities, which is much less effective than the strategy that optimizes for routing distance.

\setlength{\tabcolsep}{4.5pt}
\begin{table}[h!]
\centering
\label{tab:citiesInf}
\begin{threeparttable}
\begin{tabular}{llllllll}
\toprule
 & & \textbf{Boston} & \textbf{LA} & \textbf{NYC} & \textbf{Doha} & \textbf{Dubai} & \textbf{Riyadh} \\
\midrule
\multicolumn{2}{l}{\# facilities [x1,000]} & 66.3 & 319.8 & 207.5 & 12.9 & 38.7 & 64.0  \\

\multicolumn{2}{l}{Population [million]} & 2.36 & 9.59 & 8.12 & 0.99 & 2.91 & 5.17 \\

\multicolumn{2}{l}{Area [km$^2$]} & 1,649 & 3,869 & 1,132 & 309 & 1,402 & 1,091 \\

\multicolumn{2}{l}{$N_{occ}$} & 1,125 & 3,696 & 947 & 217 & 657 & 847 \\

\multicolumn{2}{l}{$N_{max}$} & 1,947 & 5,505 & 1,357 & 380 & 1,635 & 1,305 \\

\multicolumn{2}{l}{$UCI$} & 0.21 & 0.17 & 0.19 & 0.24 & 0.35 & 0.26 \\

\midrule
\multirow{10}{*}{{\rotatebox{90}{$D_{occ}$}}} & Hospital  & 0.11 & 0.13 & 0.27 & 0.27 & 0.20 & 0.33 \\
& School & 0.32 & 0.33 & 0.74 & 0.37 & 0.18 & 0.29 \\
& Supermarket & 0.06 & 0.08 & 0.38 & 0.13 & 0.12 & 0.14 \\
& Park & 0.42 & 0.45 & 0.74 & 0.22 & 0.17 & 0.26 \\
& Pharmacy & 0.23 & 0.30 & 0.67 & 0.22 & 0.24 & 0.37 \\
& Bank & 0.37 & 0.35 & 0.67 & 0.42 & 0.33 & 0.47 \\
& Fire Station & 0.13 & 0.09 & 0.32 & 0.03 & 0.02 & 0.01 \\
& Concert Hall & 0.05 & 0.07 & 0.14 & 0.05 & 0.03 & 0.07 \\
& Soccer Field & 0.08 & 0.04 & 0.11 & 0.14 & 0.07 & 0.10 \\
& Bar & 0.30 & 0.26 & 0.66 & 0.18 & 0.15 & 0.19 \\
 \bottomrule
\end{tabular}
\end{threeparttable}
\end{table}
\noindent {\bf Table 1. The statistical information of the six cities.}

\subsection*{Optimal Distribution of Facilities to Maximize Overall Accessibility}

Accessibility indicates the level of service of facilities to the residents. In network science, accessibility is defined as the ease of reaching points of interest within a given cost budget~\cite{antunes2003accessibility,barthelemy2011spatial,levinson2012network}. How to allocate the facilities to maximize the overall accessibility in cities is one of the most essential concerns of facility planning. From this point of view, we redistribute the facilities by minimizing the total routing distance of population to their nearest facilities. In the following, we refer to this redistribution as the \textit{optimal} scenario. Likewise, the empirical distribution of facilities is referred to as the \textit{actual} scenario. Specifically, among the $N_{max}$ blocks of one city, we denote as \textit{facility-tagged} the $N$ blocks that are occupied by a given type of facility in the \textit{actual} scenario and redistribute the same number of facilities in the \textit{optimal} scenario. The shortest distance between any pair of two blocks is calculated using the Dijkstra’s algorithm in the road network. The idea is to find a new set of $N$ blocks and label them as \textit{facility-tagged} such as it minimizes the total population-weighted travel distance from all $N_{max}$ blocks to the newly selected $N$ blocks. This optimal allocation problem in networks is known as the \textit{p-median problem} and here it is solved with an efficient algorithm proposed by Resende and Werneck~\cite{resende2007fast} (\textit{Materials and Methods}).

The difference of the travel distance between the \textit{actual} and the \textit{optimal} scenarios assesses the quality of the distribution, and therefore, of the accessibility in different cities. In each scenario, each residential block is associated with the facility that can be reached in the shortest routing distance. The block is linked to itself if it is occupied by a facility. It is important to note that we do not consider in the present study the capacity of facilities as a constraint, i.e., the number of people using the same facility is not limited. We group the set of blocks served by the same facility, and define them as a \textit{service community}. Considering hospitals as an example, we present in Figs.~2A and 2B the \textit{service communities} in Boston in the \textit{actual} and \textit{optimal} scenarios, respectively. The color of each cluster depicts the total population $p_j^S$ in the \textit{service community} of the $j$th facility. The communities in \textit{optimal} scenario are more uniform in both size and population compared to those in the \textit{actual} scenario. Particularly in the \textit{actual} scenario, the communities have small area in downtown Boston but large in the rural area, revealing the uneven distribution of hospitals.


\begin{figure*}[tb!]
\centering
\includegraphics[width=1.0\linewidth]{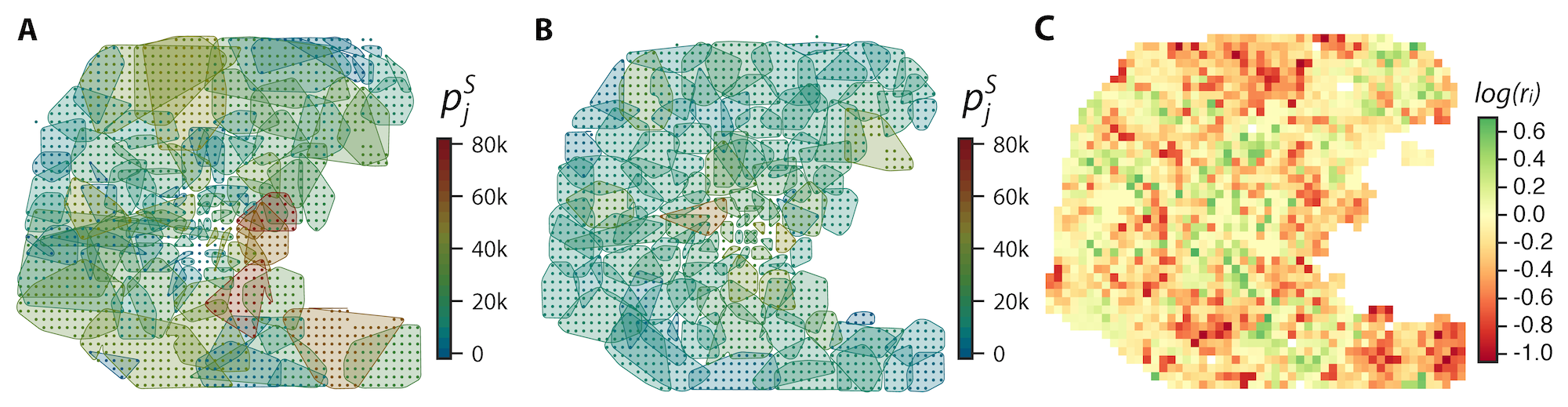}
\caption{{\bf \textit{Service communities} of hospitals in the \textit{actual} and \textit{optimal} scenarios in Boston and gain of travel distance per block.} (\textbf{A}) \textit{Service communities} of the hospitals as measured from empirical data in Boston. The dots refer to the blocks in the city. The color indicates the population in each community. (\textbf{B}) \textit{Service communities} of the optimally distributed hospitals in Boston. In this \textit{optimal} scenario, both of the area and the population of communities are generally more equable than those in the \textit{actual} scenario. (\textbf{C}) The block gain index $r_i$ in logarithmic scale in Boston. The blocks in green indicate that the residents are better served in reality, note that $r_i$ does not have units. The blocks in red indicate their actual travel distance to the hospital is larger than the optimal distance and they are underserved, such as in the northern and southeastern areas.}
\end{figure*}

In order to quantify the disparities between blocks in the level of service for a given type of facility, we compare the actual and optimal travel distances to facilities. We define a gain index of the $i$th block as:
\begin{equation}
r_i = \frac{l_i}{\hat{l}_i}\quad ,
\end{equation}
where $\hat{l}_i$ and ${l}_i$ are the shortest travel distances from the $i$th block to its nearest facility in the \textit{actual} and \textit{optimal} scenarios, respectively. A $r_i$$>$$1$ identifies that the block is better served by the facility in the \textit{actual} than the \textit{optimal} scenario. Residents living in these blocks benefit more from the distribution of facilities than they would in the scenario of social optimum. In Fig.~2C, we illustrate in Boston the $r_i$ of each block to hospitals in logarithmic scale. The blocks in green, near to hospitals, are located in the central, southern and northeastern areas, while the blocks in red have lower accessibility to hospitals when compared with the \textit{optimal} scenario and are located in the northern, southwestern, and southeastern areas. This has some resemblance with the spatial distribution of wealth in Boston metropolitan area~\cite{Census}. The actual travel distance $\hat{l}_i$ and the gain index $r_i$ in the $i$th block to hospitals for six cities are presented in fig.~S3.

Although the inequality of the distribution of facilities can be visually observed from Fig.~2C, for comparing the inequality across facility types and between cities, we compute the Gini coefficient of $r_i$ of all blocks per facility type per city, as illustrated in fig.~S4A. We observe the Gini coefficients of all selected facility types in Boston are similar and around $0.5$. NYC has the most discrepancies in the Gini coefficients over the ten facility types, where the distributions of schools, parks, pharmacies, banks and bars are more equitable than others due to their high densities (see Table~1). In the GCC cities, fire stations are the most equitably distributed facilities, while bars, hospitals, parks and pharmacies are distributed less equitably than others. The Lorenz curves and the values of the Gini coefficients per facility type are presented in fig.~S4B. The three cities in the U.S. are generally planned more equally than the GCC cities.

Thereafter, we compare the difference in accessibility across cities to various facility types. Fig.~3A presents the average travel distances in the \textit{actual} scenario ($\hat{L}$) and \textit{optimal} scenario ($L$) to the ten selected types of amenities. The first row displays the facilities with higher densities in the U.S. cities: banks, pharmacies, schools, parks and bars. Next comes hospitals and supermarkets, followed by concert halls, soccer fields and fire stations which have the lowest densities. As expected, the lower the density the longer the travel distance to them. Note that the accessibility to parks, fire stations and bars have the largest differences between U.S. and GCC cities, mainly due to lower availability in the later. To compare the travel distance in different cities in the same order, we exhibit the scatterplots of $\hat{L}$ and $L$ versus $D_{occ}$, the ratio between $N$ and $N_{occ}$, in Figs.~3B and 3C, respectively. The discrepancy of actual travel distance $\hat{L}$ among the six cities is mainly caused by the difference in facility planning strategy and urban form. As expected, the optimal travel distance $L$ displays a more uniform tendency than $\hat{L}$, revealing the potential of modeling $L$ with the number of facilities $N$.

\begin{figure*}[tb!]
\centering
\includegraphics[width=0.95\linewidth]{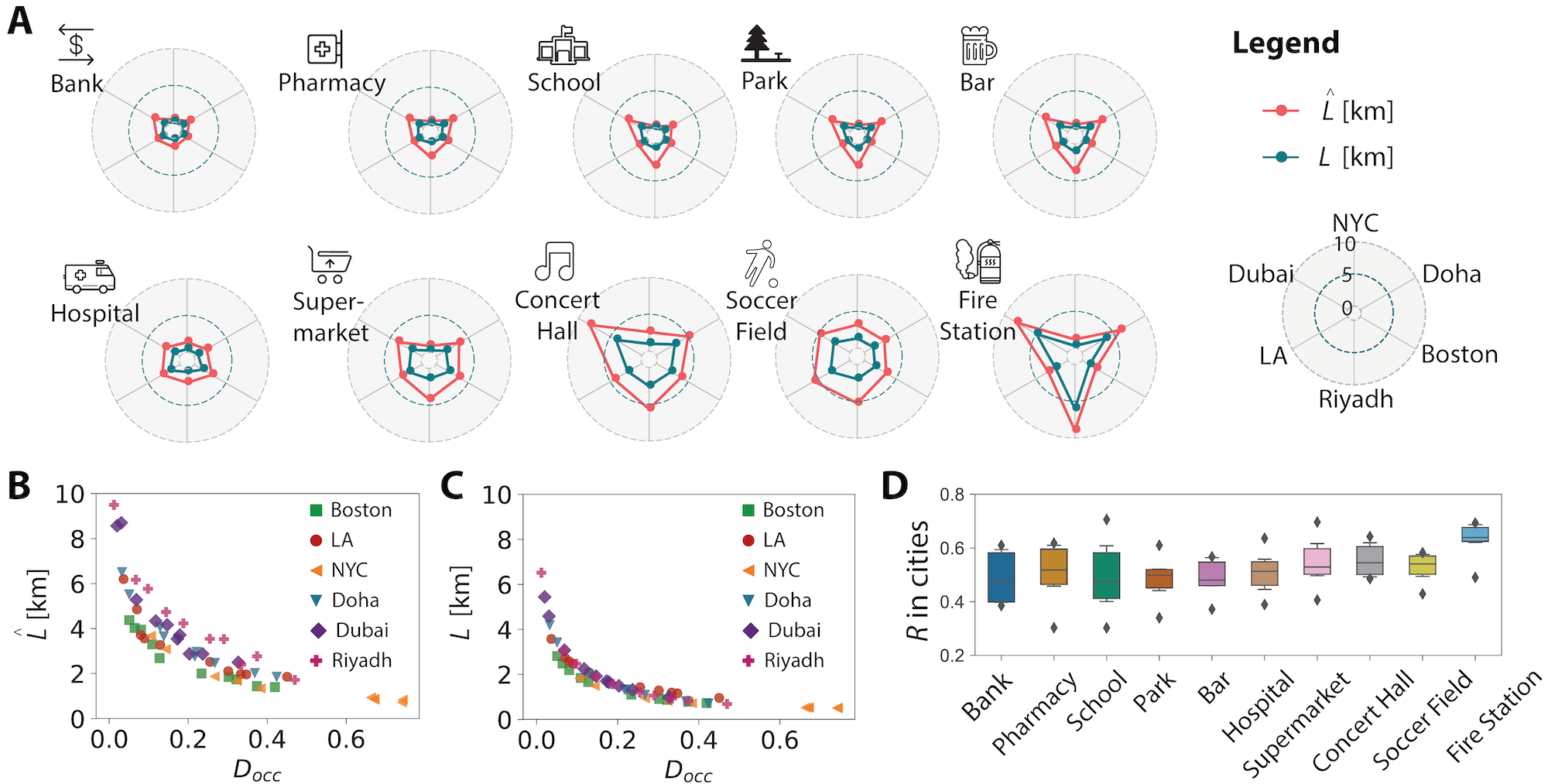}
\caption{{\bf Optimality of planning by facility type and city.} (\textbf{A}) The average travel distance in \textit{actual} scenario, $\hat{L}$, and \textit{optimal} scenario, $L$, for the ten types of facilities in the six cities. (\textbf{B}) $\hat{L}$ as a function of $D_{occ}$, the ratio between $N$ and $N_{occ}$, in the six cities. The dot refers to one of the ten types of facilities in a given city. Cities display different descending rates. (\textbf{C}) $L$ as a function of $D_{occ}$ in the six cities. All cities show clearly similar descending rates. (\textbf{D}) Box plot of the optimality index, $R$, by facility type. Facility types are ranked by their average densities in the six cities in the descending order. Among the facility types, fire station is the most optimally distributed and bank and school are the worst. In general, facility type with lower density is better located than dense facility type from the perspective of collective benefit maximization.}
\end{figure*}

An interesting measure is the improvement of overall accessibility if the locations of facilities are optimally redistributed at city scale. To that end, we define the optimality index $R$ for a given type of facility at city level as the ratio between the average travel distance to the nearest facilities in the \textit{optimal} and \textit{actual} scenarios,
\begin{equation}
R = \frac{L}{\hat{L}} = \frac{\sum_{i=1}^{N_{max}}{p_i l_i}}{\sum_{i=1}^{N_{max}}{p_i \hat{l}_i}} \quad ,
\end{equation}
where $p_i$ is the population in the $i$th block. $R$ ranges from 0 to 1, with 1 indicating the facilities are optimally distributed in reality. In note~S3 and fig.~S4C, we discuss the change of $R$ with $N/N_{max}$ by introducing two extreme planning strategies, \textit{random} and \textit{population-weighted} assignments, described in note~S3. We observe that $R$ score of actual planning is mostly between the two extreme strategies, except Riyadh, in which $R$ is even lower than \textit{random} assignment. This suggests the imbalance between facility locations and service delivery in Riyadh~\cite{alhomaidhi2019geographic}. Besides, we observe the $R$ score of actual planning is the highest when $N/N_{max}$ is the smallest for cities, except LA and NYC. For the two extreme strategies, we observe $R$ is u-shaped as a function of $N$, except for LA. This suggests higher $R$ for both small and large $N$ values. This is because for small $N$, simply allocating the facilities in the most crowded blocks would shorten the total travel cost to a great extent, while for large $N$ most blocks are occupied by facilities. The $R$ score of LA keeps flat compared to other cities mainly due to the polycentric distribution of population, indicating that a small number of facilities can not efficiently serve most of the population.

Fig.~3D depicts the box plot of $R$ of the ten types of facilities in the six cities. $R$ is generally lower with larger facility density, suggesting the gaps between \textit{actual} and \textit{optimal} distribution are larger. For example, hospitals and fire stations have much lower density than bars but their $R$ scores are larger. Public services need to be uniformly distributed while commercial ones do not. The more available facilities are banks, pharmacies, schools, parks and bars, with $R$ between $0.4$ and $0.5$ on average, revealing that the average travel distance could be reduced by $50\%$ if all facilities are planned in the optimal locations.

\subsection*{Revisiting Scaling Law between Facility and Population Densities}

\begin{figure*}[htb!]
\centering
\includegraphics[width=0.95\linewidth]{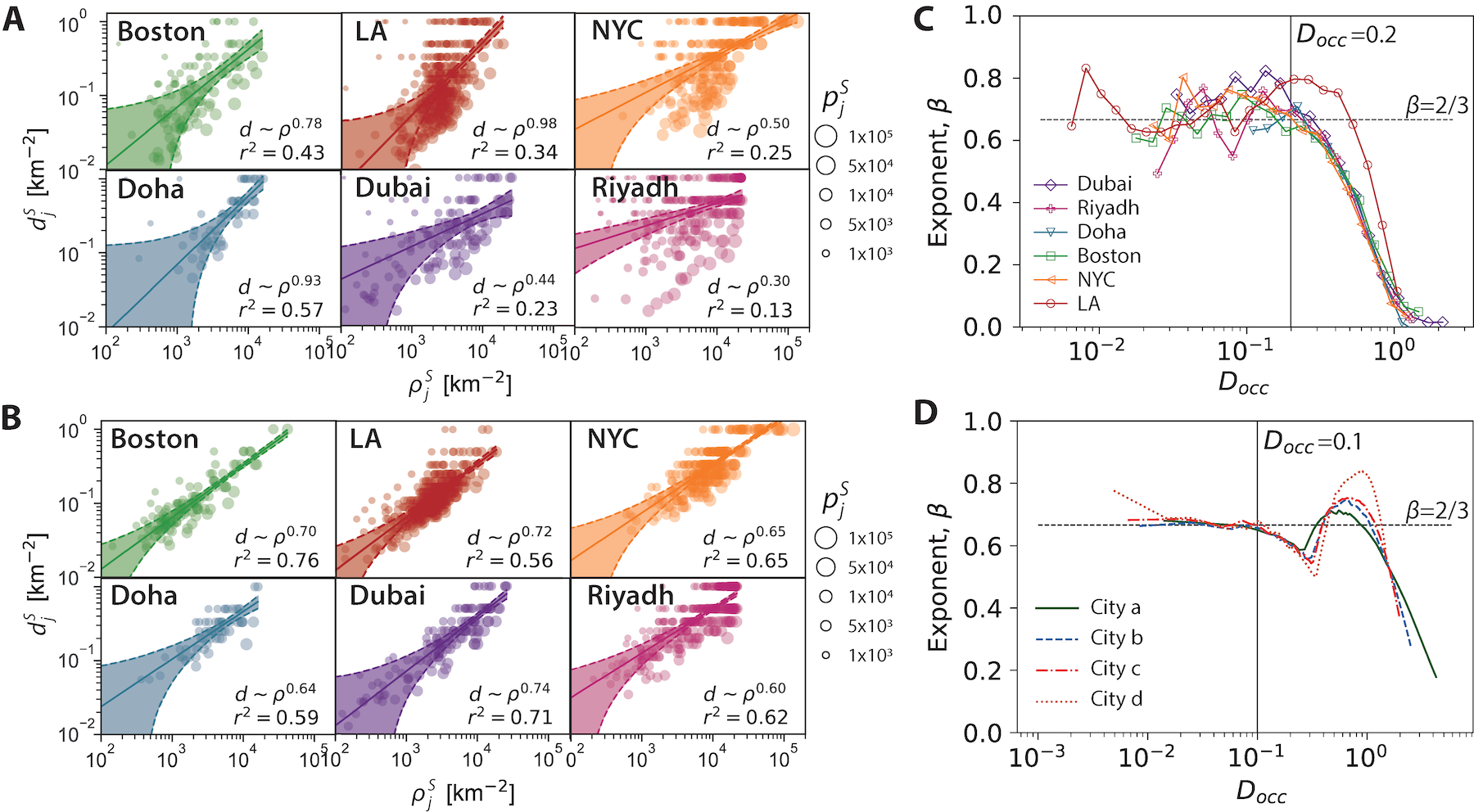}
\caption{{\bf Fitted power law for the distribution of hospitals per city.} (\textbf{A}) Actual hospital density (inverse of the area of community) vs. population density at \textit{service community} level. Each colored dot refers to a \textit{service community} and the size represents its population, as shown in Fig.~2A. The full line represents the best fitted power law. The colored shadow represents the $95\%$ confidence interval. The low $r^2$ indicates that a clear power law can not be found and the fitted exponent differs from the empirically observed $2/3$ at larger scales~\cite{um2009scaling}. (\textbf{B}) Hospital density vs. population density in the \textit{optimal} scenario. The full lines show clear power laws, with exponents close to $2/3$ in all cities. The $r^2$ are higher and confidence intervals are narrower than the \textit{actual} scenario. The fitted exponents for other facility types are numerically provided in table~S1. (\textbf{C}) Change of fitted exponent $\beta$ with $D_{occ}$ in the \textit{optimal} scenario in the six cities. Each $\beta$ is calculated by optimally distributing a given number of facilities in the city. The gray dashed and full lines indicate $\beta$$=$$2/3$ and $D_{occ}$$=$$0.2$, respectively. (\textbf{D}) Change of fitted exponent $\beta$ with $D_{occ}$ in the \textit{optimal} scenario in four toy cities.}
\end{figure*}

Previous work has related the facility density to population density as a power function both in the \textit{actual} and \textit{optimal} scenarios~\cite{um2009scaling} at the national scale. Here, through introducing the road networks, we dissect these power laws in the two scenarios in diverse cities. We calculate both facility and population densities in the \textit{service communities}, as shown in Figs.~2A and 2B. Specifically, $d_j^S$$=$$1/a_j^S$, and $\rho_{j}^S$$=$$p_j^S/a_j^S$, where $a_j^S$ is approximated by the product of the number of blocks $n_j^S$ and the average block area in the city, that is $a_j^S$$=$$n_j^S \bar{a}$. Taking hospitals as an example, their densities versus the population densities of the \textit{service communities} in the \textit{actual} scenario over the six cities are illustrated in Fig.~4A. The full lines represent the fitted power law functions with least squares method and with communities with more than $500$ residents. Cities have different exponents and the $r^2$ scores of the fitting are less than $0.5$ in most cases. These results show that, despite the $2/3$ power law was found for public facilities at county-resolution~\cite{um2009scaling}, we do not find a uniform law between facility and population densities at finer resolutions i.e., intra-city community level.

Once facilities are optimally redistributed in the city, the \textit{service communities} are reorganized accordingly. The fitted power laws between the distribution of hospitals and population in \textit{optimal} scenario of the six cities are shown in Fig.~4B. The fitted exponents are closer to $2/3$ and have larger $r^2$, and the $95\%$ confidence intervals are narrower than those in Fig.~4A, depicting the \textit{actual} scenario. The exponents for the ten selected types of facilities in the \textit{actual} and \textit{optimal} scenarios are reported in table~S1. As expected, cities have different exponents for both \textit{actual} and \textit{optimal} scenarios. In all cases, we observe that the optimal exponents deviate from the analytical $2/3$ previously reported when the facilities are optimally distributed by Euclidean distance at national case~\cite{gastner2006optimal}. Sources of difference are both the constraints introduced by the road networks and the higher density of facilities to be distributed.

For a comprehensive understanding of the existence of the power laws, we optimally allocate varying number of facilities $N$ in our six cities of study and in synthetic cities. In Figs.~4C and 4D, we relate the $\beta$ to $D_{occ}$, the ratio of $N$ to $N_{occ}$, and observe $2/3$ when $D_{occ}$$<$$0.2(0.1)$ for the real-world (synthetic) cities. We simulate controlled scenarios via four synthetic or toy cities of size 100$\times$100, with population distributions depicted in Fig.~5A. Note that the population threshold is set as 50 in toy cities to count $N_{occ}$, and the total population is fixed as half million, which is about $1/10$ of the studied cities. We find the curves of diverse cities collapse into a single one, indicating the difference in the change of $\beta$ across cities is mainly caused by different $N_{occ}$. Interestingly, in the toy cities, we notice that the change of $\beta$ is not monotonous. It stays around $2/3$ when $D_{occ}$ is below $0.1$. Subsequently, $\beta$ decreases with $D_{occ}$ as more facilities are assigned to the low density regions and then increases as facilities start to refill the high density regions. After all high density blocks are assigned with facilities, $\beta$ starts to drop to zero, implying all blocks are filled with facilities. The same fluctuation of $\beta$ is not clearly observed in real-world cities because the large and low density regions are not segregated like in the synthetic cities. In summary, in the \textit{optimal} scenario, the $2/3$ power law can be found for a limited number of facilities, but tends to disappear for larger values of $N$.

\subsection*{Modeling Accessibility to Optimally Distributed Facilities}

\begin{figure*}[htb!]
\centering
\includegraphics[width=0.95\linewidth]{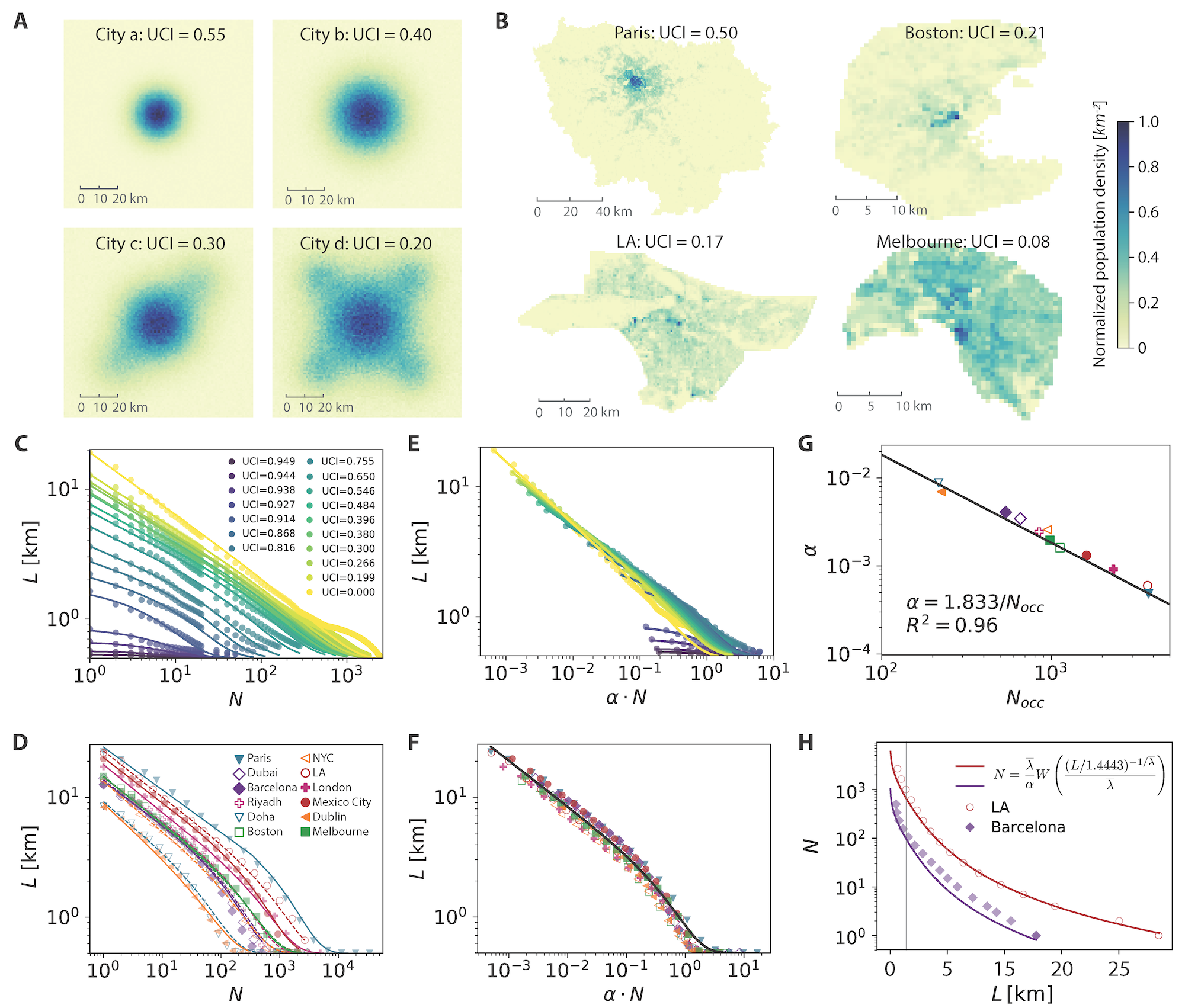}
\caption{{\bf Modeling of optimal travel distance to facilities in toy and real-world cities.}
(\textbf{A}) Population distributions of four selected toy cities. (\textbf{B}) Population distributions of four selected real-world cities, among which Paris is the most centralized and Melbourne is the most polycentric. 
(\textbf{C-D}) Simulated and modeled optimal travel distance, $L$, versus the number of facilities, $N$, in toy cities and real-world cities, respectively. The dots represent the simulated $L$ with varying $N$. The lines represent the fitted model $L(N)$ in Eq.~\ref{eq:finalL}. Cities are ranked by UCI in the descending order in the legend.
(\textbf{E}) Simulated and modeled $L$ versus $\alpha N$ in toy cities. By scaling $N$ with $\alpha$, we collapse the curves of $L$ with UCI lower than 0.9 into a single one.
(\textbf{F}) Simulated and modeled $L$ versus $\alpha N$ in real-world cities. The black line represents the function in Eq.~\ref{eq:finalL_uni}, approached by the simulated $L$ in all cities. 
(\textbf{G}) Relation between $\alpha$ and $N_{occ}$. $\alpha$ can be well fitted with $N_{occ}$, suggesting the decay rate of shared population in blocks without facilities is in inverse proportion to the urban area in one city for a small $N$.
(\textbf{H}) Validation of the universal function in LA and Barcelona.}
\end{figure*}

In Fig.~3B we see that $D_{occ}$ is the most determinant factor to decrease the average distance $\hat{L}$ to a facility independent of its type and city. Interestingly, in Fig.~3C we observe that these decreasing functions $L(D_{occ})$ collapse for the optimal distributions in each city. Following up on this observation, we explore further the relation between travel distance in \textit{optimal} scenario $L$ and the number of facilities $N$ for diverse cities with various geographic constraints and population distributions. To this end, we design $17$ toy cities with different levels of urban centrality, four of these illustrated in Fig.~5A. Population distributions of the toy cities are generated by a two dimensional Gaussian function (e.g., cities a and b) or a mixture of several two dimensional Gaussian functions (e.g., cities c and d). The toy cities have the same population of half million and are equal-sized, consisting of 100$\times$100 blocks. The size of each block is set to 1 km$^2$ and travel cost between two blocks is calculated with the Euclidean distance between their centroids. We measure the centrality of a city by computing the urban centrality index (UCI), proposed by Pereira et~al.~\cite{pereira2013urban}, of the population distribution (\textit{Materials and Methods}). UCI ranges from 0 to 1, with 0 indicating the totally polycentric --- with the population of the city uniformly distributed, and 1 indicating totally monocentric --- with all the population residing in one block. In addition, we include 12 real-world cities for further exploration, the six aforementioned to which we add: Paris, Barcelona, London, Dublin, Mexico City, and Melbourne. Population distributions of four selected cities are illustrated in Fig.~5B. Paris is the most monocentric with UCI of $0.50$ and most residents residing in the urban region, while Melbourne is the most polycentric with UCI of $0.08$ and residents dispersed over the city.

For an estimate of the optimal travel distance $L$ in each city, we first assume the in-block travel distance is constant $l_{min}$$=$$0.5$ km, and the average travel distance within a \textit{service community} approximates to $g^S_j\sqrt{a^S_{j,occ}}$, where $g^S_j$ denotes the geometric factor in the community; $a^S_{j,occ}$ denotes the area of the occupied blocks~\cite{gastner2006optimal}. Then $L$ is expressed as the sum of two terms, the first for the population in the $N$ blocks with facilities and the second for the population in the $N_{max}$$-$$N$ blocks without facilities:
\begin{equation}
L=\frac{1}{P} \cdot \left( l_{min} \cdot \sum_{j=1}^N{p_j} + \sum^{N}_{j=1} g^S_j {\tilde{p}}^S_{j}(a^S_{j,occ})^{0.5}\right) \quad,
\label{eq:timeTwoTerms}
\end{equation}
where $P$ is the total population in the city, ${\tilde{p}}^S_j$ denotes the population in the \textit{service community} of the $j$th facility after removing the block where the $j$th facility is located, that is ${\tilde{p}}^S_j$$=$$p^S_j$$-$$p_j$. We find that $a^S_{j,occ}$ follows power law relation to the total area in community $a^S_j$ in most cities, that is $a^S_{j,occ} \propto (a^S_j)^\gamma$ (fig.~S5A). We assume that $g^S_{j}$ is constant in each city, written as $g_{city}$, and $a^S_j \approx \overline{a^S} = \bar{a} \cdot N_{max}/N$ with $\bar{a}$ denoting the average block area in the city. Then we can rewrite Eq.~\ref{eq:timeTwoTerms} as
\begin{equation}
    L(N)=l_{min} \cdot p(N) + A\cdot N^{-\lambda}\cdot (1-p(N)) \quad ,
\end{equation}
where $p(N)$ denotes the share of population in blocks with facilities; $A$ and $\lambda$ are both constant. More details of this derivation can be found in note~S4.1.

We further study how the share of population in blocks without facilities is related to the number of facilities $N$, and find $1-p(N)\approx e^{-\alpha N}$ when $N \ll N_{occ}$ (see details in fig.~S5B, notes~S4.2 and S4.3). Thereby, we could model $L$ as
\begin{equation}
    L(N) = l_{min} \cdot \left(1 - e^{-\alpha N}\right) + A\cdot N^{-\lambda} \cdot e^{-\alpha N} \quad ,
\label{eq:finalL}
\end{equation}
where the number of facilities $N$ is the main variable that determines $L$. While $\alpha$ controls the relation between $p(N)$ and $N$; $A$ and $\lambda$ are two free parameters to calibrate. The model of $L(N)$ summarizes the fact that to model $L$ the only two essential ingredients are the number of facilities to allocate $N$ and the distribution of population in space. 

Next, we numerically assign the optimal distribution of facilities given varying number of facilities for both toy and real-world cities. We present the average travel distance $L$ versus the number of facilities $N$ in the toy and real-world cities in log-log plots in Figs.~5C and 5D, respectively. In Fig.~5C, we see that for the same $N$, the global travel costs in polycentric cities are larger than the monocentric ones. To validate the proposed function $L(N)$, we first calibrate $\alpha$ by fitting $1-p(N)=e^{-\alpha N}$ per city (fig.~S5B), then calibrate the two free parameters A and $\lambda$ in Eq.~\ref{eq:finalL} with the simulated $L$. All parameters are presented in table~S2. The fitted $L(N)$ are shown with lines in Figs.~5C and 5D. The simulated and modeled $L$ are presented separately for each city in fig.~S6, showing good results in various empirical conditions.

For seeking a universal function to approach the simulated $L$ in diverse cities, we use $\lambda$ in Eq.~\ref{eq:finalL} as a constant, fixing its average empirical value $\bar{\lambda}=0.382$ in the $12$ real-world cities. Combining the observation that $N_{max}$ is inversely proportional to $\alpha$ and $A \approx g_{city} \overline{a}^{\overline{\lambda}} N_{max}^{\overline{\lambda}}$ (note~S4.1), we can expect that $A\propto \alpha^{-\overline{\lambda}}$. Figure S7C confirms this, showing that $A=1.4443\alpha^{-\overline{\lambda}}$. We can rewrite Eq.~\ref{eq:finalL} as follows: 
\begin{equation}
    L(N) = l_{min} \cdot \left(1 - e^{-\alpha N}\right) + 1.4443 \cdot (\alpha N)^{-\overline{\lambda}} \cdot e^{-\alpha N} \quad ,
\label{eq:finalL_uni}
\end{equation}
This function with only one free parameter $\alpha$ suggests that we are able to rescale $N$ with $\alpha$ to collapse the curves of $L$ in all cities into one, as shown in Fig.~5F that depicts Eq.~\ref{eq:finalL_uni} as solid line. The same rescaling of $N$ in toy cities is presented in Fig.~5E, where the collapse is not as good as in the real cities due to the divergent values of $\lambda$ of toy cities in table S2. Next, we go beyond the average distance $L$ and plot the distribution of travel distances when keeping $\alpha N$ fixed (figs. S7F and S7G). In all cases the travel distance follows a Gamma distribution. This universality suggests that: (i) given a certain $\alpha N$, all real-world cities can reach comparable accessibility; (ii) the overall accessibility in the \textit{optimal} scenario not only depends on the availability of the resources but also the settlement of population, independently from the road network and total area of the city.

Empirically, the decay of population share in blocks without facilities $\alpha$ depends on the population distribution in space. Taking into account that unpopulated blocks are not ideal when optimizing accessibility, $N_{occ}$ is a better variable to express $\alpha$. A good agreement $\alpha$$=$$1.833/N_{occ}$ ($R^2$$=$$0.96$) over the 12 real-world cities is shown in Fig.~5G, suggesting that $\alpha$ can be estimated by $N_{occ}$. Given that $\alpha$$=$$1.833/N_{occ}$ and the universal relation of $L(\alpha N)$, we can explain the collapses found in Figs.~3C, 4C and 4D.

As a concrete application of this universal model for optimal distance of facilities, in Eq.~\ref{eq:finalL_uni}, we can plan for facilities by, for example, extracting how many facilities are needed for varying levels of accessibility to a given type of service. In this context, the number of facilities $N$ can be estimated with the inverse function of Eq.~\ref{eq:finalL_uni}. As the second term in Eq.~\ref{eq:finalL_uni} dominates the $L$ for a limited $N$, we simply invert the second term to estimate $N$, given by $N(L;\alpha) = \frac{\bar{\lambda}}{\alpha} \cdot W(\frac{(L/1.4443)^{-1/\bar{\lambda}}}{\bar{\lambda}})$, where $W(\cdot)$ is the ProductLog or Lamber-W function (note~S4.5). Fig.~5H presents the estimated and simulated $N$ versus $L$ for two limiting cases, LA, in which the approximation agrees well with the simulation, and Barcelona, in which the approximation underestimates $N$. The results of other real-world cities are depicted in fig.~S8, showing in general a good agreement between the analytical approximation via the Lamber-W and the numerical simulations.

\section*{Discussion}

As cities differ in their form, economy and population distribution, the inter-play between population and facility distributions are challenging to plan. The accessibility of facilities are constrained by their availability, the road network and means of transportation. While efforts are devoted to managing daily commuting and transit oriented developments, the planning of the distribution of different urban facilities deserves attention to a paradigm shift towards walkable cities. We present a framework that uses publicly available data to compare the optimal and the actual accessibility of various facility types at the resolution of urban blocks. This allows us to efficiently pinpoint blocks that are under-served, i.e. those where people have to travel longer distances to reach the facilities they need compared to the social optimum. By relocating the facilities to optimize the global travel distance, we find that the relation between facility and population densities follows the scaling law, $d$$\propto $$\rho^\beta$ only in the limit of few or limited number of facilities, regardless of the differences in road network structures. The observed exponent $\beta$ is generally around $2/3$ if the number of facilities is diluted or less than $10\%$ of the occupied blocks, and it starts to decay for larger number of facilities. This confirms the continuous limit for diluted number of facilities presented at national scale~\cite{um2009scaling}. We observe that the empirical conditions within cities do not follow the continuous approximation for the power law with population density, because facilities are not equally planned and the number of facilities is large in comparison with the number of populated blocks.

To gain further insights when the number of facilities is large, we analytically model the average travel distance $L$ in the \textit{optimal} scenario vs. the number of facilities $N$ and three parameters. Parameter $\alpha$ represents the rate of the population share in blocks without facilities, and the other two parameters can be approximated as constant among cities. A universal expression $L(\alpha N)$ is verified with 17 synthetic cities and 12 real-world cities depicting diverse urban forms. Furthermore, the travel distance to optimally distributed facilities follows a Gamma distribution for all cities once $\alpha N$ is fixed. This function can be applied to estimate the number of facilities needed to offer services to people within a given accessibility in average. The results estimated with the derived function find a good match to the numerical simulations that require solving the optimal distribution of facilities. When relating $\alpha$ to the urban form, we uncover that centralized cities require less facilities than polycentric cities to achieve the same levels of accessibility. Applications of this framework could be to optimally reallocate resources that provide emergency services, such as the placement of shelters, ambulances, or mobile petrol stations in the event of natural disasters. 

The optimal planning of facilities in this work supposes that all residents equally need the resources and the accessibility is measured from their places of residence. In reality, the socio-economic segregation in cities results in heterogeneous needs for resources. Cities in different social systems and economic development levels also exhibit different needs for various types of facilities that would need to be taken into account for economic considerations. On the other hand, people’s needs are naturally dynamic and change in time and space owing to their time-varying mobility behavior. All these factors result in complex interactions between the allocation of facilities and settlements of residents, and can be considerable avenues for future research. Another important avenue is to consider the limited capacity of facilities in the optimal planning. This became ever more evident when distributing the healthcare system resources during the outbreak of a pandemic, such as the COVID-19 in 2020.

\section*{Materials and Methods}
\subsection*{Datasets description}

\begin{enumerate}
\item[] {\bf Population density} The population with a spatial resolution of 30 arc-seconds (approximately 1 km$^2$ near the equator) of each city was obtained from the LandScan~\cite{landscan2015} in 2015. The average population density varies from 1,431 per km$^2$ in Boston to 7,175 per km$^2$ in NYC.

\item[] {\bf Facility data} Facilities were crawled from Foursquare using their public APIs in 2017~\cite{foursquare2017}. In the dataset, each facility is associated with a name, geographical location, a facility type (e.g., hospital, supermarket, bank) and a category (e.g., Arts \& Entertainment, Nightlife Spot, Shop \& Service). Facilities are then assigned to the blocks defined by the LandScan population data with their locations. The total number of facilities in the six cities are given in Table~1. The distributions of facility categories for each city are presented in fig.~S1. We select 10 types of facilities to inspect their actual and optimal planning and they are given in the legend of Fig.~1. After merging facilities of the same type located in the same block, the occupancy of each type of facility, $N/N_{occ}$, are presented in Table~1. The U.S. cities generally have more dense facilities than the GCC cities.

\item[] {\bf Road networks} We extract the road networks from OpenStreetMap~\cite{osm2017}. The road network is represented as a directed graph, in which edges indicate road segments and nodes indicate intersections. Each edge is associated with a weight representing its length. The travel distance between two blocks is computed by finding the shortest path between two randomly selected nodes in these blocks using Dijkstra algorithm~\cite{dijkstra1959note}.
\end{enumerate}

\subsection*{Optimal Distribution of Facilities in Space}
Finding the optimal locations of facilities to minimize the total travel cost is essentially an optimal placement problem in network theory, which is NP-hard and known as \textit{p-median problem}. The problem in this work is formalized as follows: ``\textit{Given a set of blocks $N_{max}$ in a city, a set of residential blocks $X \in N_{max}$ are with population, and each block in $N_{max}$ can only accommodate one facility. The goal is to open $N$ facilities in $N_{max}$ so as to minimize the sum of population-weighted travel distances from each residential block to its nearest open facility.}''~\cite{dohan2015k}

For simplicity, the \textit{p-median problem} is written as a linear programming problem.

\begin{equation}
\begin{aligned}
& \text{minimize} &   & \sum_{i}{\sum_{j}{c_{i,j} x_{i,j}}} & & \\
& \text{subject to} & & \sum_j{x_{i,j}}=1 & & \forall i \\
& & & \sum_{j}y_j=N & & \\ 
& & & x_{i,j} \leq y_j & &  \forall i, j\\
& & & x_{i,j} \in \{0,1\} & & \forall i, j \\
& & & y_j \in \{0,1\} & & \forall j
\end{aligned}
\end{equation}
\vspace{5mm}

\noindent where $i$ and $j$ are indices of the blocks; $x_{i,j}$$=$$1$ means that people living in block $i$ are assigned to their nearest facility in block $j$, and $i$$=$$j$ signifies that there is a facility located in residential block $i$; $y_j$$=$$1$ if there is a facility in block $j$, else $y_j$$=$$0$; $N$ is the number of facilities to assign and we assume that one block can only accommodate one facility of the same type; $c_{i,j}$ is the travel cost from block $i$ to block $j$, which equals to the total routing distance of all population residing in block $i$. In this work, we solve the \textit{p-median problem} with a fast algorithm based on swap-based local search procedure implemented by Resende and Werneck~\cite{resende2007fast}.

\subsection*{Urban Centrality Index}

We adopt the UCI proposed by Pereira et~al. to measure the centrality of the population distribution in cities~\cite{pereira2013urban}. UCI is the product of two components, the location coefficient (LC) and the proximity index (PI). The former is introduced to measure the inhomogeneity of population distribution in space. The latter is introduced to measure the difference between the current distribution and the most decentralized scenario. The calculation of LC and PI are as follows.
\begin{equation}
    LC = \frac{1}{2}\sum_1^{N_{max}}{\left(s_i - \frac{1}{N_{max}}\right)}
\end{equation}
\begin{equation}
    PI = 1 - \frac{V}{V_{max}}
\end{equation}
where $V = \mathcal{S}' \times \mathcal{D} \times \mathcal{S}$. $\mathcal{S}$ is a vector of population fraction in block $i$, $s_i$$=$$p_i/P$, signifying the share of population in block $i$ ($p_i$) of the total population of the city ($P$); $\mathcal{D}$ is the distance matrix between blocks. $V_{max}$ is calculated by assuming the total population are uniformly settling on the boundary of the city, which indicates an extreme sprawl. UCI ranges from 0 to 1. Large UCI values indicate more centralized population distributions.


\bibliographystyle{Science}

\section*{Acknowledgments}
This work was supported by the QCRI-CSAIL, the Berkeley DeepDrive (BDD) and the University of California Institute of Transportation Studies (UC ITS) research grants. 

\section*{Author contributions}
YX, LEO, SA and MCG conceived the research and designed the analyses. YX and SA collected the data. YX and LEO performed the analyses. MCG and YX wrote the paper. MCG supervised the research.

\section*{Competing interests}
The authors declare that they have no competing interests. 

\section*{Data and materials availability} All data needed to evaluate the conclusions in the paper are present in the paper and/or the Supplementary Materials. Additional data related to this paper may be requested from the authors.

 
\section*{Supplementary materials}
Note S1: List of variables and notations for the accessibility analysis.\\
Note S2: Data description.\\
Note S3: On the optimality index.\\
Note S4: Derivation of the average optimal travel distance.\\
Fig. S1. Road network, population density, and the facilities in each category in the six cities.\\
Fig. S2. Distribution of population, facilities, and travel distance in the six cities.\\
Fig. S3. Actual travel distance and gain index to hospitals of each block.\\
Fig. S4. Gini coefficient of the block gain index and optimality index per facility type per city.\\
Fig. S5. Fitting the number of occupied blocks and the share of population in blocks without facilities.\\
Fig. S6. The modeled and simulated average travel distance in optimal scenario in toy and real-world cities.\\
Fig. S7. Simple model of travel distance with number of facilities.\\
Fig. S8. Comparison between the simulation results and the universal function.\\
Table S1. Best fitted exponent of the power law for the actual and optimal distribution of facilities.\\
Table S2. Fitted parameters of the 17 toy cities and the 12 real cities.\\
References ~\cite{battyFractalCities,Alonso,Mills,Muth,Clark}.    


\end{document}